# Modulation of trion and exciton formation in monolayer $WS_2$ by dielectric and substrate engineering

Tamaghna Chowdhury[1‡], Diptabrata Paul[1‡], Divya Nechiyil[1], Gokul M. A[1], Kenji Watanabe[2], Takashi Taniguchi[3], G.V. Pavan Kumar[1] and Atikur Rahman*[1]

[1] Department of Physics, Indian Institute of Science Education and Research, Pune, Maharashtra, India, 411008.
[2] Research Center for Functional Materials, National Institute for Material Science, Tsukuba, Ibaraki 305-0044, Japan.
[3] International Center for Materials Nanoarchitectonics, National Institute for Material Science, Tsukuba, Ibaraki 305-0044, Japan.

‡ These authors contributed equally

E-mail: atikur@iiserpune.ac.in

## Abstract

Photoluminescence (PL) of transition metal dichalcogenide (TMD) monolayers is strongly influenced by the dielectric environment. The defect states present in the substrate induces uncontrollable doping in the TMD monolayer and thereby modifies the PL spectra. There have been enormous efforts to tune and overcome the effect of inevitable subtract defects in PL spectra, but a proper understanding and a convenient way are still lacking. Here, we systematically studied the effect of surface defects by gradually increasing the separation between $WS_2$ monolayer and substrate. Hence, we could precisely modulate the exciton and trion contribution in the PL spectra of $WS_2$. The excitation power dependant measurements on dielectric engineered and patterned substrates helped us to shed light on the mechanism of PL modulation in monolayer $WS_2$. We have also studied the influence of the nature of the charge carried by substrate defects on the PL spectra. These results open a new pathway to modulate and obtain the desired PL spectra of TMDs by engineering the substrates. Our findings will be useful for fabricating excitonic interconnects, valleytronic, and single-photon devices.

# 1. Introduction

Transition metal dichalcogenides (TMDs) have attracted considerable interest in recent times due to their unique electrical and optical properties [1–15]. These TMDs are usually grown in their monolayer form by chemical vapor deposition (CVD) [16–19] or are mechanically exfoliated [20] from their bulk form. TMD monolayers have high carrier mobility, high absorption coefficient which can be harnessed in making optoelectronic devices such as light emitting diode (LED) photodetectors, high-performance optoelectronics [8,21–24]. Unlike its bulk form, monolayer TMDs exhibit transition from an indirect to direct bandgap [9, 21, 25–27]. Additionally, due to the reduced dielectric screening (compared to their bulk form), optical transition phenomena in the monolayer TMDs are highly susceptible [28–30] to the surrounding dielectric medium which includes the underlying substrate as well [31–35]. Inefficient screening of Coulomb interactions between the photoexcited carriers favours the formation of excitons and charged excitons (trions) in these 2D semiconducting materials [36, 37]. The signature of excitons and trions in the optical transition is strongly influenced by the nature of the underlying substrate and the surrounding medium on the top [35, 38]. Usually, thermally grown $SiO_2$ on silicon is used as a substrate for the monolayer TMDs but the TMD/$SiO_2$ interface may contain many charge impurities which act as a local doping centre. These doping centres affect the nature of optical transitions like photoluminescence (PL) to a large extent [39, 40]. It is reported that the effect of these doping centres can be reduced by placing the monolayer TMDs on hBN [35, 40, and 41]. The doping centres usually contribute extra carriers (electrons or holes) in the TMD channel and help in trion formation [42]. Thus, by tuning the dielectric environment, one can minimize the effect of doping centres at the TMD/$SiO_2$ interface, hence can modulate the contribution of trions and excitons in the PL spectra. There are reports which investigated how various substrates (silicon, hBN, mica, quartz, etc.) affect the PL emission of TMDs like $WS_2$ [35, 38, 43–47]. It has been theoretically predicted that if we increase the spacing between substrate defects and the monolayer $WS_2$, the neutral exciton starts dominating gradually with respect to the trion. However, the lack of systematic experimental studies [48, 49] motivated us to examine the PL modulation of monolayer TMDs either by inserting different thicknesses of hBN as a spacer layer or by engineering the dielectric environment. The present study sheds light on the mechanism of PL modulation by controlling the effect of impurity via dielectric and substrate engineering and provides a recipe for getting desired PL from monolayer TMDs.

Here we have varied the local dielectric environment and tuned the doping at the interface by taking substrates of varying materials and textures. CVD grown $WS_2$ monolayer has been taken as a model 2D system and 300 nm thermally grown $SiO_2$ on Si and hBN (exfoliated on $SiO_2$/Si) of varying thickness have been used as substrates, which were patterned with micron-size holes or pillars for further dielectric modulations. First, we have placed a CVD-grown monolayer $WS_2$ on top of hBN (which is exfoliated on a $SiO_2$/Si substrate) of varying thickness to reveal the effect of impurities on the PL spectra of $WS_2$. When the $WS_2$ was on $SiO_2$/Si the PL spectra predominantly showed a single trion peak. If $WS_2$ is placed on a very thin (~6 nm) hBN layer, both the trion and exciton peaks were observed. With increasing thickness of hBN layer, the trion peak started diminishing and when $WS_2$ is placed on a very thick ($\geq$ 90 nm) hBN, only exciton peak was present. $WS_2$ monolayers were also placed on a hole and pillar patterned $SiO_2$/Si substrate to understand how a gap (air) between $WS_2$ and $SiO_2$ is effective in isolating the doping centres. By varying the excitation laser power in each of these cases, we studied how exciton and trion contribution in PL evolves with increasing number of photogenerated carriers inside the $WS_2$ channel. We observed that the evolution of the ratio of trion to exciton peak intensity as a function of excitation power in $WS_2$ on top of hBN is markedly different from that of a suspended $WS_2$ in the hole region. This illustrates that these two different methods of isolating the doping centres have different effects on the PL of $WS_2$. We have further placed an hBN layer on top of a $WS_2$ layer which was partly placed on a bottom hBN layer (the remaining part of this $WS_2$ is on $SiO_2$) in such a way that one part of the $WS_2$ is encapsulated between two hBN layers and the other part has an hBN layer just on top while bottom part was on $SiO_2$. This was done to understand the contribution of doping centres at the $WS_2$/$SiO_2$ interface and surroundings on the PL of $WS_2$. Finally, by modifying the sign of the charge on the substrate surface, we examined how the polarity (+ve or -ve) of the surface defects affects the PL of $WS_2$. Our study elucidates the effect of impurity on the PL of $WS_2$ and provides a reliable way to generate excitons and trions selectively, which will have a profound impact on exciton interconnects [50], valleytronic devices [13–15], and single-photon emission processes from the defects of 2D materials [51].

# 2. Results and Discussions

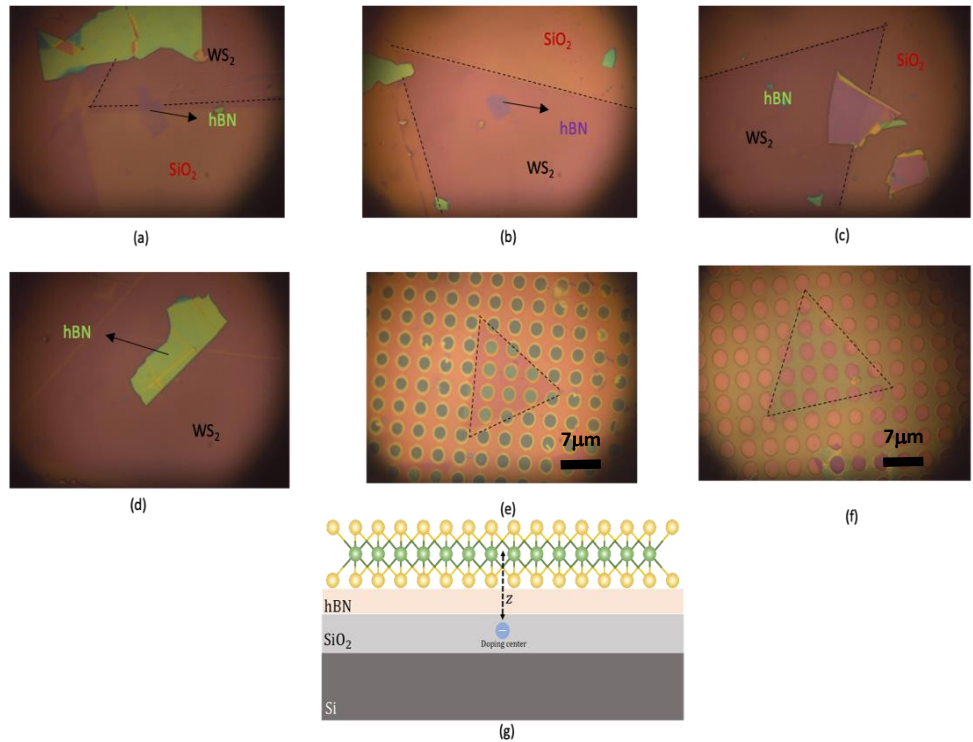

**Figure 1.** Optical microscope image (at 100X) of WS$_2$ monolayer on (a) thinnest (~ 6 nm) hBN (b) slightly thicker (~ 9 nm) hBN (c) thicker (~ 90 nm) hBN (d) a very thick (~200 nm) hBN. (e) WS$_2$ on a hole patterned SiO2/Si substrate (highlighted by dashed black line) and (f) WS$_2$ on a pillar patterned SiO$_2$ substrate (highlighted by black line) (g) Schematic showing role of hBN as spacer.

Monolayer WS$_2$ was synthesized using a home-built atmospheric pressure chemical vapour deposition (APCVD) unit (see methods for details). The monolayer WS$_2$ flakes were transferred on a 300 nm SiO$_2$/Si substrate containing hBN flakes of varying thicknesses (see supplementary material figure 1 and Table 1) using polystyrene (PS) film [52] (figure 1a-d). To study the optical properties of suspended monolayer WS$_2$, it was transferred on a hole patterned SiO$_2$/Si substrate (figure 1e). These holes have a diameter of approximately 4 $\mu m$, depth of 190 nm, and inter hole separation (centre to centre) of 6.5 $\mu m$ (see supplementary figure 2a). These parameters were chosen such that their dimensions are always greater than the laser spot size (~ 2 μm) to avoid any averaging effect in our optical measurement data. Another kind of sample was made which had pillars of SiO$_2$/Si having diameter 4 $\mu m$, height 190 nm, and interpillar separation (centre to centre) of 6.5 $\mu m$. (figure 1f) (see supplementary material figure 2b). For both hole and pillar structures, patterns were made by photolithography, and etching was done using reactive ion etching (RIE). WS$_2$ monolayers were transferred on the hole and pillar patterned substrates by the PS transfer technique. These textured substrates were used to minimize the effect of substrate-induced defects since the contact area with SiO$_2$ will be minimum as part of the WS$_2$ monolayer will remain suspended. To encapsulate WS$_2$ using hBN to make hBN/WS$_2$/hBN heterostructure, we used the dry transfer technique [53] (see methods for details).

First, we started with investigating PL spectra of a monolayer WS$_2$ on a 300 nm SiO$_2$/Si substrate. The PL spectra exhibited a single peak at ~1.95 $eV$ (figure 2a). It has been reported that the PL peak of a transferred WS$_2$ monolayer is slightly redshifted (almost 50 meV) compared to that of an as-grown WS$_2$ monolayer. This is due to the relaxation of strain when the WS$_2$ layer is transferred [54]. There is another report according to which PL peak of transferred WS$_2$ gets blue-shifted compared to the as-grown one because of the trapped water layer in the transferred sample, which acts as a source of doping [38]. We took PL on our as-grown WS$_2$ monolayer and the shift was less than 10 meV compared to the transferred one (see supplementary figure 3). When monolayer WS$_2$ was placed atop a thin (~ 6 nm) hBN layer (figure 1a, marked as hBN 1), PL spectra (figure 2a) showed a prominent peak at ~1.96 $eV$ with another small feature at ~2.00 $eV$. The low energy peak is attributed to negatively charged trion (X$^-$) contribution while the higher energy peak is arising due to neutral exciton (X$^0$) contribution [55,56] . We compared the PL spectra with another monolayer WS$_2$ placed on a slightly thicker hBN of 9 nm thickness (figure 1b, marked as hBN 2). Here we observed that the X$^-$ peak at ~1.96 eV becomes less prominent while X$^0$ peak at ~2.01 eV becomes sharper and more prominent (figure 2a). The study was repeated by increasing the bottom hBN thickness, which led to the observation that with increasing thickness of hBN, the X$^-$ peak (figure 2a, marked as hBN 3) gradually becomes less prominent while for a very thick bottom hBN (~90 nm for figure 1c and ~200 nm for figure 1d) the X$^-$ peak vanishes and the only prominent peak is X$^0$ at ~2.00 eV (figure 2a, marked as bulk hBN). Thus, we observed that we are getting neutral exciton signature in PL just by placing the monolayer WS$_2$ on hBN flake which was otherwise dominated by trions when the WS$_2$ layer was placed directly on SiO$_2$/Si substrate (figure 2a). In other words, the trion to exciton peak intensity ratio decreases with an increase in thickness of hBN layers (see supplementary figure 4). This is in accordance with the theoretical works of Tuan *et. al.* [57], which predicted that with the increase in separation of monolayer WS$_2$ from the surface defects or impurity (situated on SiO$_2$) the relative trion to exciton intensity ratio decreases. A normalized form of the plot in figure 2a is provided in the supplementary information (see supplementary figure 5).

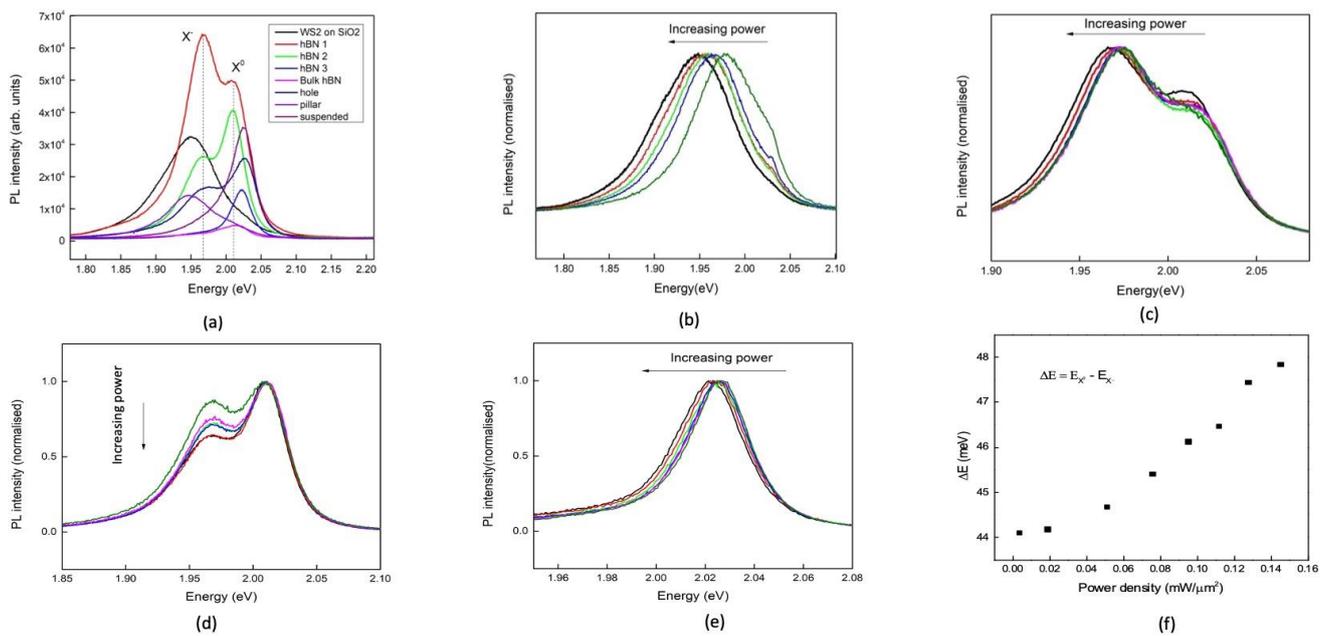

**Figure 2.** (a) PL spectra of various cases at 0.394 mW/μm² excitation power density showing exciton and trion peaks. Excitation power density (0.002 to 0.394 mW/μm²) dependent PL spectra (normalized) of WS$_2$ on (b) SiO$_2$/Si substrate (c) 6 nm hBN (d) 9 nm hBN (e) 90 nm hBN (f) Energy difference of exciton and trion ( ΔE) as a function of excitation power density.

The modulation of X$^-$ and X$^0$ peaks with hBN thickness can be attributed to the presence of the unintentional source of doping in the SiO$_2$/WS$_2$ interface [55]. This source of doping is the charged defects present on the SiO$_2$ surface. In presence of doping, the WS$_2$ PL is dominated by the trions. Formations of trions are more favourable when there is more number of carriers inside the WS$_2$ channel. The unintentional doping sources are providing these additional carriers inside the channel. Because of that, we are seeing a broad X$^-$ peak in the PL of WS$_2$ placed directly on SiO$_2$/Si [54]. In other cases, by placing the WS$_2$ layer on a very thin hBN flake, the effect of the doping source gets screened up to some extent and we can see well-resolved X$^-$ and X$^0$ peaks. A little shift of ~5 $meV$ can be due to the flake-to-flake variation of PL. This shift is not due to the strain effect, because Raman spectra of all these samples are similar (supplementary figure 6a) [58]. Now, as we increase the thickness of the hBN flake, the effect of the doping source diminishes further due to dielectric screening by hBN and also due to increasing spatial separation between SiO$_2$ and WS$_2$. Thus, for a WS$_2$ on thick (≥ 90 nm) hBN, the spectra are dominated by neutral exciton peak X$^0$.

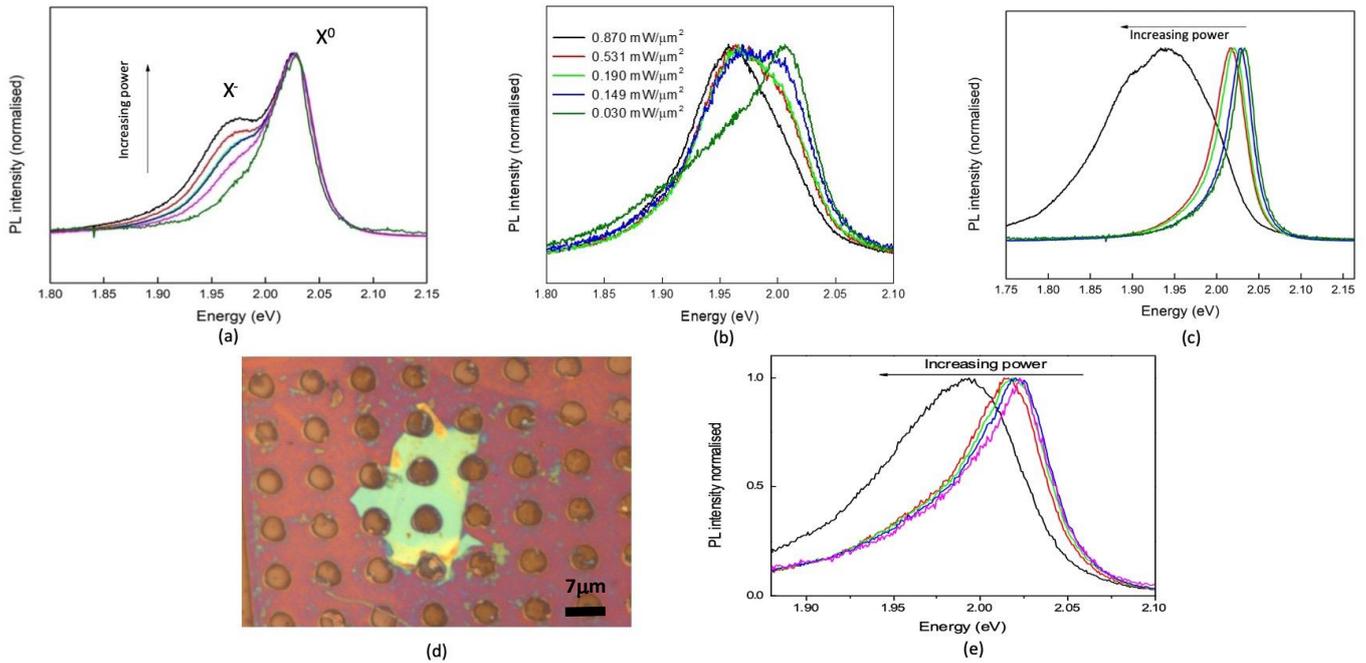

**Figure 3.** Excitation power dependent (0.03 to 0.870 mW/μm²) PL spectra (normalized) of WS$_2$ on (a) suspended region on hole patterned SiO$_2$/Si substrate (b) pillar (WS$_2$ in contact with SiO$_2$) (c) the suspended WS$_2$ region of pillar patterned SiO$_2$/Si substrate. (d) Optical microscope image (100X) of monolayer WS$_2$ on hole patterned hBN substrate. (e) suspended region on hole patterned hBN substrate (shown in figure 3d).

We have performed laser power-dependent PL for all of these samples to understand the exact mechanism of PL modulation due to dielectric engineering. For the WS$_2$ on SiO$_2$/Si, the power-dependent PL (figure 2b) shows a clear redshift of the $X^-$ PL peak with increasing power. This indicates that the peak is dominated by trions and there is less contribution of neutral excitons. This is because with increasing laser power the carrier density increases, which raises the dissociation energy of trions causing the redshift [39, 56, 59]. Now, in the laser power dependant PL for WS$_2$ on a very thin (6 nm) hBN (figure 2c) we see that the $X^-$ peak has a slight redshift with increasing power but less compared to the previous case of WS$_2$ on SiO$_2$/Si. The $X^0$ peaks show almost no observable shift with increasing excitation power which is consistent with the literature [39, 56, 60]. This less shift in trion peak in WS$_2$ on hBN case compared to WS$_2$ on SiO$_2$/Si can be attributed to the diminishing effect of unintentional sources of doping from the SiO$_2$/WS$_2$ interface. In figure 2f, energy difference of trion and exciton in figure 2c is plotted as a function of excitation power. The increase of energy separation with increasing laser power indicates an increase in carrier density in the channel [61–63]. Since, a relatively higher excitation power was used in our experiment, peak intensity of raw PL spectra with laser power density was plotted (supplementary figure 7(a)-(c)) to see any signature of Auger effect [64]. But from the analysis (see supplementary information section 7 for more details) of the data no signature of Auger effect was observed and PL is primarily dominated by free carrier and excitonic recombination. Probably, the excitation power used in our experiments are not high enough to see a Auger recombination in WS$_2$ since this happens only at very high excitation power densities [65]. In the case of WS$_2$ on SiO$_2$, photogenerated carriers from unintentional doping centres as well as from donors of the n-doped channel were contributing to the PL. When hBN is introduced between SiO$_2$ and WS$_2$, the effect of unintentional doping centres is minimized to some extent but not completely [40] causing lesser redshift of $X^-$ peaks. This suggests that a thin hBN layer screens the effect of charged impurities up to a certain extent at the substrate interface. Excitation power dependant PL of WS$_2$ on a slightly thicker (9 nm) hBN (figure 2d) shows that with increasing power, trion peak intensity decreases. The power dependant PL of WS$_2$ on a bulk (90 nm thick) hBN (figure 2e) shows a small redshift of exciton peak with no observable trion peak at any laser power. The small redshift of the exciton peak in figure 2e can be due to the laser induced heating of the sample. Although, in the excitation power dependent Raman measurement (supplementary figure 8) no detectable change in the Raman peaks were observed [66, 67]. But it is important to mention that with our measurement setup a maximum resolution of 1.15 cm$^{-1}$ in wavenumber can be achieved, whereas for monolayer WS$_2$, rate of change of Raman shift has been reported to be -0.0091cm$^{-1}$/K (For E"$_{2g}$) [68]. Therefore, at least a temperature difference of ~126 K is required to be detectable in our measurements (further details are included in the PL and Raman spectroscopy of methods section). Based on our observation we can say that if there is any heating effect, the maximum temperature of our WS$_2$ samples is less than ~419 K (considering ambient temperature to be 293K). However, the possibility of heating due to laser where the temperature can rise even by few tens of degrees, which can cause such redshift, [69] cannot be ruled out.  For better understanding, peak width and position as a function of excitation power for trion and exciton in figure 2c are shown in the supplementary figure 9a and 9b respectively. The decrease of intensity of trion with the increase of excitation power in the case of WS$_2$ on top of 9 nm hBN is described while explaining figure 5a in the latter half of this paper. To further study the effect of substrate, we transferred monolayer WS$_2$ on hole and pillar patterned SiO$_2$/Si substrate. We took PL spectra on the hole region (figure 2a) where the WS$_2$ layer is suspended freely (figure 1e). The PL spectra exhibit similar characteristics to that of a WS$_2$ on moderately thick (9 nm) hBN but with a slight blueshift in X$^0$ and X$^-$ peaks. This shift is probably due to the strain imparted on the suspended region inside the hole which is evident from the Raman spectra (supplementary figure 6b) [58]. We studied the excitation power dependant spectra on the hole region (figure 3a) and observed redshift in the X$^-$ peaks and comparatively less redshift in the X$^0$ peaks (peak width and position as a function of laser power is plotted in supplementary figure 9c and 9d). In contrast to the PL spectra of WS$_2$ on thin (6 nm) hBN (see figure 2c), in this case, PL spectra at the lowest excitation power exhibited minimal X$^-$ contribution. Probably this is because, at the same excitation power, in the WS$_2$ on hBN case, photon flux was able to ionize the required number of donors so that there is trion formation. On the other hand, in the case of suspended WS$_2$, the number of donor atoms is less as the WS$_2$ is freely suspended and the unintentional doping centres at the SiO$_2$/WS$_2$ interface are absent. Thus, the same photon flux was unable to ionize the required number of donors for trion formation and hence trion peak is absent in the PL spectra. But as we increase the laser power, the increase in photon flux can ionize sufficient donor atoms and we see X$^-$ peak emerging at higher excitation powers. Whereas, in bulk (90 nm) hBN (figure 2e), even at the highest excitation power no X$^-$ peak was observed. This suggests that bulk hBN is more efficient in screening the surface defects. In the case of WS$_2$ on hBN, hBN acts as a dielectric medium while for a hole suspended WS$_2$, air trapped inside the hole acts as a dielectric medium. Since hBN has a higher dielectric constant of ~3.5 compared to air and because of its large thickness (~90 nm), electric field lines from the surface defects are screened more effectively [60]. Probably, this is the reason why we don't see a trion peak in the case of WS$_2$ on top of 90 nm hBN. In the case of hole suspended WS$_2$, there is a finite volume of air trapped inside the hole which is not the case in WS$_2$ placed on thick hBN. O$_2$ present in the trapped air may get physisorbed in WS$_2$ and this can deplete the electron concentration inside the channel [70, 71]. The adsorption energy of O$_2$ in TMDs typically ranges between 0.1~1.9 eV [72, 73]. Thus photons of energy 2.33 eV are sufficient to break this bond. But at low intensity, there are not that many photons available to desorb significant amount of O$_2$ and we only see the excitonic feature. As we increase the laser intensity, probably more and more O$_2$ get desorbed from the channel and electron concentration starts to rise. Thus at higher excitation power, a trion like feature can be seen [74, 75]. The PL spectra (also excitation power dependence) taken on WS$_2$ on SiO$_2$ (non-hole region) showed only X$^-$ feature similar to that of the earlier case of WS$_2$ on SiO$_2$. In case of monolayer WS$_2$ on pillar patterned substrate (figure 1f), the PL spectra taken on pillars (figure 2a) (i.e., WS$_2$ on SiO$_2$) showed only trion peak while the spectra on the suspended region showed only the exciton peak (peak width and position as a function of laser power is plotted in supplementary figure 9e and 9f for better clarity). In the excitation power dependant spectra of the pillar top region (figure 3b) we observed that at the lowest power, the exciton peak is dominant. As the laser power is increased, the peak gradually shifts to the trion side and at the highest power, only the trion peak is visible. From the field effect scanning electron microscope (FESEM) image (supplementary figure 10) of the pillar patterned substrate, we found that the monolayer WS$_2$ sits conformally on the pillar top which has a diameter of around 4μm while the laser spot used had a size of 2μm. Raman spectra were taken on the pillar top and suspended region away from the pillar top at the same excitation power (supplementary information figure 11). From Raman spectra, it was observed that E''$_{2g}$ peak is redshifted by ~ 4.7 cm$^{-1}$ in pillar top compared to the suspended region or flat SiO$_2$/Si substrate. This observation is a well-known signature of tensile strain in monolayer WS$_2$ where the Raman peak (especially E''$_{2g}$) gets redshifted with increasing strain [76,77]. Our results show that strain is more on the pillar top region compared to the suspended region. It is also known from the literature that, strain gradient in monolayer TMDs create a force on neutral exciton and pushes them towards maximum strained region i.e. the region where band gap becomes minimum due to strain [78,79]. Probably, in our case, because of strain, exciton formation becomes more favourable in the pillar top region which is not the case when WS$_2$ is just placed on a flat SiO$_2$/Si substrate. But as we increase the laser power, photo-induced carrier number increases, and now a neutral exciton is more likely to bound with another electron to form trion. Thus, in the case of $WS_2$ on pillar top, we see a gradual transition from exciton to trion as we go from lower to higher power. The excitation power dependant PL (figure 3c) on the area where $WS_2$ is suspended showed only $X^0$ peak and no $X^-$ peak (except at the highest excitation power) compared to $WS_2$ in the hole region. To further investigate the effect of the surrounding donor centres on a suspended $WS_2$ monolayer, PL spectra were studied on a sample in which monolayer $WS_2$ was placed on holes made on an hBN flake exfoliated on a $SiO_2$/Si substrate (figure 3d). In this study, on the hole pattern (of hBN) we observed only $X^0$ peak and no $X^-$ peak (figure 3e). So, the surrounding donor centres cannot affect the PL in this case and trion peak was not observed at any excitation power. This also explains the observation of only exciton peak in the case of suspended $WS_2$ on a pillar patterned substrate. In figures 3c and 3e, a huge redshift of the exciton peak at the highest power is observed. This can be attributed to the formation of electron hole plasma (EHP) at higher laser powers. We argue that this shift is probably not due to the heating effect because the observed shift in the figure is ~ 100 meV and width ~ 150 meV whereas for heating effect, shift and width are typically in the range of 40 meV and 35 meV respectively [80]. To rule out the effect of heating due to laser power we did Raman measurement (supplementary information figure 12) at different excitation powers, and we did not observe any significant change in the peak positions of $E"_{2g}$ and $A'_{1g}$ with excitation [66,67]. This suggests that, probably, heating due to laser has no significant effect on the PL spectra. The peak position, width vs. excitation power (supplementary figure 9f) were plotted and the trend is similar to that of electron hole plasma (EHP) in $MoS_2$ as was reported by Yilling yu et. al. [80]. However, this requires further investigation to confirm the presence of EHP and will be explored in our future work.

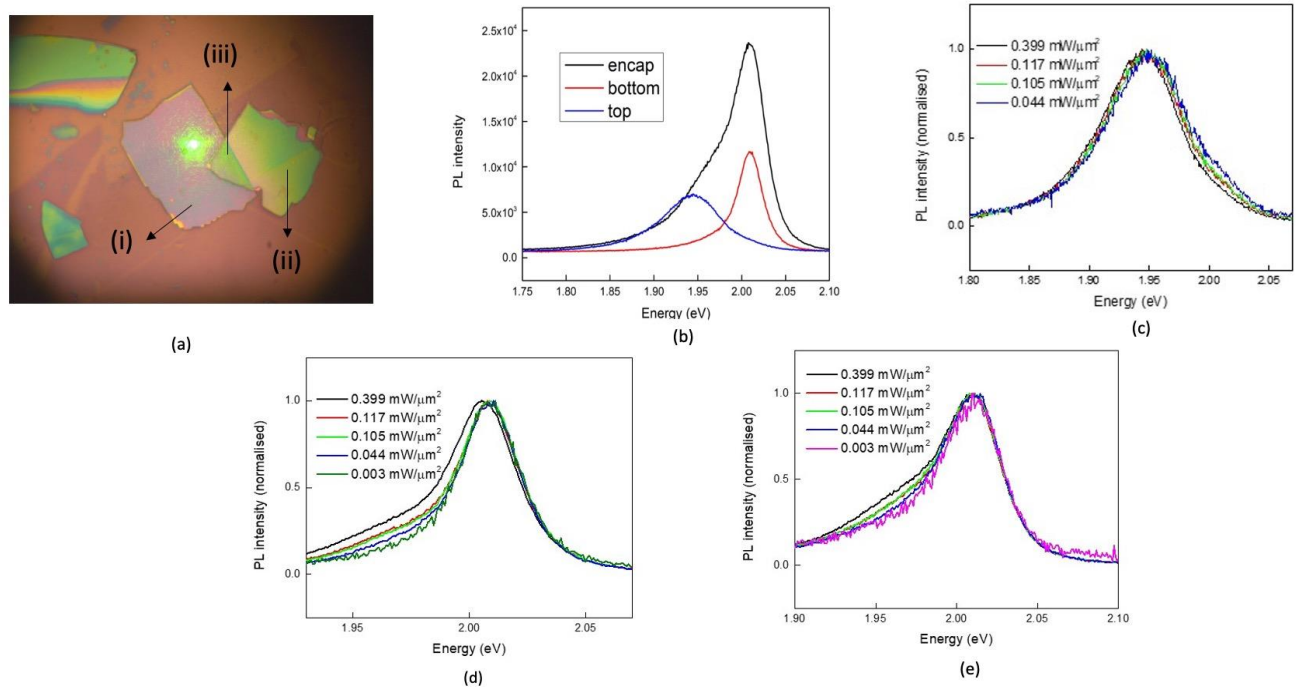

Figure 4. (a) Optical microscope image (at 100X) of i. WS$_2$ on top of hBN ii. hBN on top of WS$_2$ iii. WS$_2$ sandwiched between top and bottom hBN. (b) PL spectra from region i, ii and iii of the sample shown in Fig. 3a   Normalized PL with varying excitation power (c) on WS$_2$ with hBN on top. (d) on the WS$_2$ on top of hBN (e) on the WS$_2$ encapsulated between hBN.

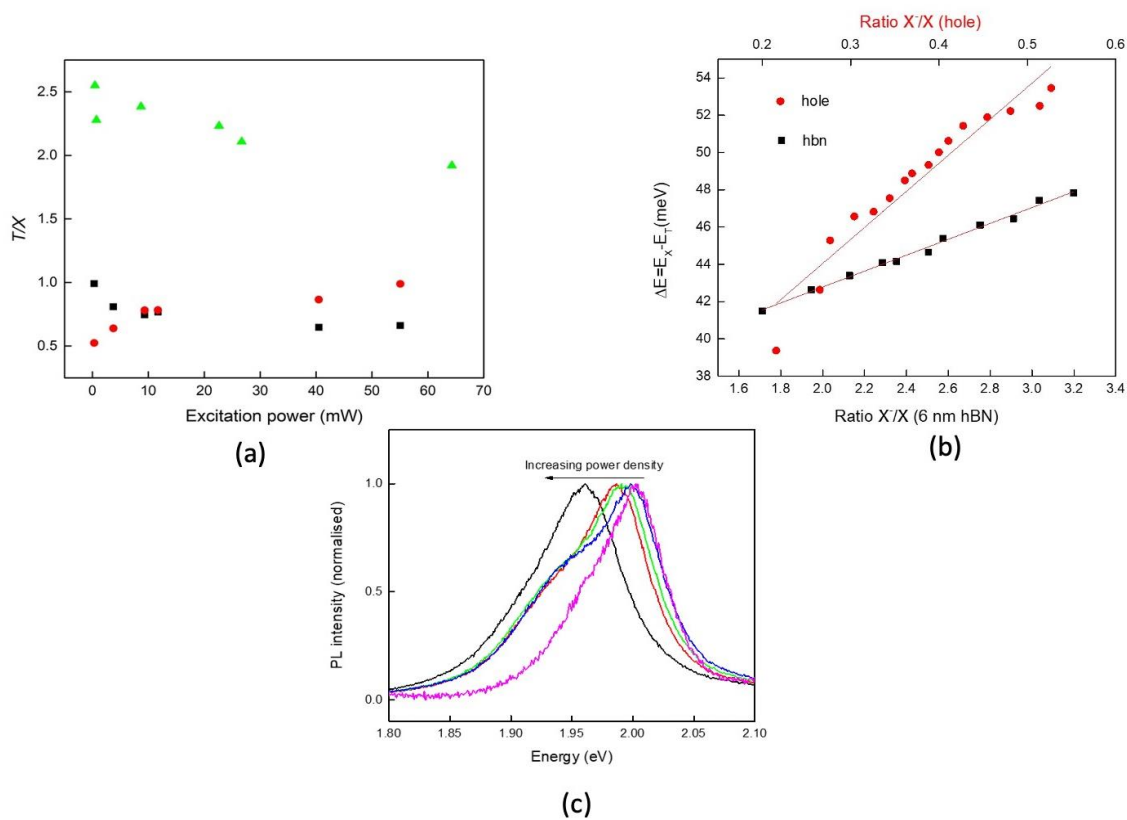

**Figure 5.** (a) Ratio of trion peak intensity and exciton peak intensity as a function of excitation power (0.03 to 0.870 mW/μm²) for i. WS$_2$ on 6 nm hBN (green triangles) ii. WS$_2$ on top of a 9 nm hBN (black squares) iii. WS$_2$ suspended on hole patterned SiO$_2$/Si substrate (red circles). (b) Splitting energy (ΔE) of exciton and trion as a function of ratio of the intensity of the exciton and trion for monolayer WS$_2$ supported on 6 nm thick hBN (black squares) and suspended on top of hole (red circle). (c) Normalized PL with varying excitation power of WS$_2$ on positively charged SiO$_2$/Si surface.

| Configuration | Trion | | Exciton | |
|---|---|---|---|---|
| | Peak energy (eV) | Width (meV) | Peak energy (eV) | Width (meV) |
| $WS_2$ on $SiO_2$/Si | 1.946 | 73.10 | ---------- | ---------- |
| $WS_2$ on 6 nm thick hBN | 1.966 | 64.14 | 2.015 | 33.64 |
| $WS_2$ on 9 nm thick hBN | 1.964 | 60.89 | 2.010 | 32.57 |
| $WS_2$ on 90 nm thick hBN | ---------- | ---------- | 2.021 | 33.69 |
| $WS_2$ suspended on hole | 1.969 | 82.41 | 2.026 | 34.49 |
| $WS_2$ on top of pillar | 1.962 | 90.11 | ---------- | ---------- |
| $WS_2$ suspended on pillar | ---------- | ---------- | 2.015 | 43.19 |
| $WS_2$ encapsulated between hBN | ---------- | ---------- | 2.008 | 63.70 |

**Table 1: Summary of the peak width and position of data shown in figure 2a**

To further support our claim that the unintentional source of doping at the $SiO_2/WS_2$ interface is affecting the PL and not any external factor, we did PL study on a monolayer $WS_2$ with different encapsulations. In region I (figure 4a), $WS_2$ is on top of hBN, one part is on $SiO_2$ but the top is covered by a hBN flake (region II) and another part of $WS_2$ is encapsulated between two hBN flakes (region III). The PL spectra (figure 4b) on region II showed only trion peak at $\sim 1.96\ eV$. By increasing excitation power, we saw only a small redshift of $X^-$ peak (figure 4c). At any excitation power, we did not observe any signature of exciton in the spectra. The PL spectra (figure 4b) of the region I, exhibit an exciton peak at around $\sim 2.01\ eV$, and a small shoulder-like feature in the trion position at around $\sim 1.96\ eV$ (figure 4d). This confirms our claim that the unintentional source of doping at the SiO$_2$/WS$_2$ interface is largely affecting the PL rather than any other factors; also, by screening these doping sources, we can tune the PL emission spectra. Now, in region III where WS$_2$ is encapsulated between two hBN flakes, we see a X$^0$ PL peak (figure 4b) at ~2.01 $eV$ and no X$^-$ peak. The excitation power-dependent spectra (figure 4e) from region I and region III exhibit similar exciton and trion characteristics. This implies that putting a top layer hBN does not have a significant impact and the impurities at the SiO$_2$/WS$_2$ interface play the most crucial role in determining the relative X$^-$ and X$^0$ contribution in PL. It has been reported that [81,82] encapsulating a TMD layer by hBN makes the PL peak narrow and sharp. But here we see that when hBN is just at the bottom, the X$^0$ peak of WS$_2$ is narrower and sharper than the encapsulated one. The PL emission is more enhanced in the case of the encapsulated region compared to other cases.

We also investigated the importance of the thickness of the bottom hBN flake for selective excitation of the X$^-$ in PL of WS$_2$. As it was seen earlier while discussing the results of figure 2c (WS$_2$ on top of 6 nm hBN), at the lowest excitation power we observed that the X$^-$ peak is dominant. Now, as we increase the excitation power, we see an X$^0$ peak emerging slowly. At the highest excitation power, we see a prominent X$^0$ peak though for all powers X$^-$ peak was more dominant and intense. To know about relative X$^0$ and X$^-$contribution we did Lorentz fitting of the PL spectra of figure 2c and 2d for deconvolution of the peaks at these powers (see supplementary figure 13). We see that with increasing excitation power, the T/X (ratio of peak intensities of X$^-$(trion) and X$^0$(exciton) respectively from the Lorentz fit) ratio decreases (figure 5a). In this case, X$^-$ peak is far more intense than the X$^0$ peak at all powers. In the case of WS$_2$ on top of 9 nm hBN (figure 2d), we see that at lowest power both X$^-$ and X$^0$ peaks of comparable intensity are present. As we increase the laser power, from Lorentz peak fit of the figure 2d, we see that the T/X ratio decreases (figure 5a). But this time at all excitation powers, X$^0$ was more intense compared to the X$^-$ peak. Now when we do the same analysis on the PL spectra of WS$_2$ suspended in the hole region (figure 3a) we see that at the lowest excitation power only X$^0$ peak is present. As we increase the laser power the X$^-$ peak starts appearing gradually. From the deconvolution of the PL spectra (figure 3a) at all these powers, we see that T/X increases with increasing excitation power (figure 5a). But the X$^0$ peak was always far more intense than the X$^-$ peak at all these powers. The T/X intensity ratio decreases as the excitation power is increased for the case when monolayer WS$_2$ is placed on 9 nm thick hBN. This is because hBN limits the number of doping centers affecting the WS$_2$ monolayer resulting in decreased excitation of trions to that of excitons as the laser power is increased. A similar effect can also be observed for the 6 nm thick hBN, although the relative intensity of the X$^-$, in this case, is always higher due to decreased screening (because of less thickness of hBN) of the doping centers. In contrast, in the case of hole-suspended monolayer WS$_2$, the decreased screening efficiency leads to an increase of trion excitation relative to that of exciton as the excitation power is increased. We have plotted the splitting between X and T as a function of their intensity ratio (figure 5b) for WS$_2$ on top of 6 nm hBN and WS$_2$ suspended on a hole patterned substrate. We observed that in both the cases (hole and hBN) splitting shows a linear dependence with T/X ratio which is expected from the earlier reports in the literature [61]. Therefore, in both the cases (hBN and hole) trion dissociation energy increases with increase of excess carriers. But the slope of the plot in case of WS$_2$ suspended on top of hole is higher than the slope of the plot in case of WS$_2$ on top of 6 nm hBN. This suggests that hBN more efficiently screens (due to the high dielectric constant of hBN) the effect of unintentional doping centres at the WS$_2$/SiO$_2$ interface compared to the hole suspended WS$_2$. This increases the T/X ratio as the excitation power is increased. As discussed earlier while explaining figure 3a, light-induced desorption of oxygen in case of holes might also contribute to this observation [70,71].

To investigate how the polarity of the defects affects the PL of WS$_2$, we have prepared SiO$_2$/Si substrate whose surface is made positively charged by coating the substrate with APTES [83] (see supplementary information section -2 for details). In the excitation power-dependent PL (figure 5c) we found that almost for every power, X$^0$ peak is the dominant one with a little hump at the X$^-$ position. At the highest power, we see a X$^-$ peak and no X$^0$ peak. This is because the positive charge of the surface neutralizes the excess electrons in the naturally n-doped WS$_2$ channel and thereby decreases the carrier density inside the channel. As a result, we observe X$^0$ peak in WS$_2$ when placed on the positively charged substrate [84]. This also means that the usual SiO$_2$/Si surface that is being used in our experiments has negatively charged surface defects which aided the trion formation in WS$_2$. We have confirmed this by treating a SiO$_2$/Si substrate with oxygen plasma (60W, 10 mins) which turns the SiO$_2$/Si surface entirely negatively charged [85]. PL spectra taken on a WS$_2$ monolayer transferred on this substrate showed only trion peak at all excitation power (data not shown).

## 3. Conclusions

In conclusion, we have demonstrated that substrate plays an important role in optical transitions like photoluminescence. The substrate and TMD interface is a major source of charged impurity which modifies the PL spectra significantly. The effect of these charged impurities or doping centres can be screened by inserting a dielectric layer like hBN and by tuning the thickness of hBN one can modify the presence of trions and excitons (see Table 1 for peak position and width of trion and exciton in different substrate configuration). In such a way we can study them together or independently which is essential for realizing valleytronic devices. We also studied the screening effect of substrate defects by suspending the WS$_2$ monolayer. Our study reveals that suspending and inserting a hBN layer have two distinct effects in the ratio of trion and exciton intensity with increasing excitation power. Also, the PL emission of the WS$_2$ monolayer strongly depends on the polarity of the charge carried by the substrate defects. These results demonstrate a facile way to modulate the PL spectra of TMDs by controlling the effect of impurities through substrate engineering. Our findings will have a profound impact in tailoring single-photon emission processes from the defects of 2D materials and fabricating excitonic interconnects and valleytronic devices.

## 4. Methods

### 4.1 WS$_2$ synthesis by APCVD

WS$_2$ monolayers were grown by APCVD on 300 nm thermally grown SiO$_2$ on Si. Properly cleaned and O$_2$ plasma-treated (60W, 5mins) substrates were placed on an alumina boat containing 500mg WO$_3$. A boat containing WO$_3$ powder was placed inside a 35mm quartz tube in the heating zone of the furnace. Another boat containing 500 mg sulphur was placed upstream inside the tube, 15 cm away from the WO$_3$ boat. The sulphur boat was outside the heating zone and was heated using a heater coil. The tube was flushed with 500 sccm Ar for 20 mins. The furnace was kept at 850°C and the heater coil was kept at 240°C for evaporation of sulphur. This set of temperatures was maintained for 10 minutes. After the growth, the system was allowed to cool naturally.

### 4.2 Preparation of hole and pillar SiO$_2$/Si substrates

Photolithography and reactive ion etching (RIE) methods were used for the substrate modification of SiO$_2$/Si substrate. First, photolithography was employed for making circular array patterns on SiO$_2$/Si substrate. Afterward, keeping photoresist as mask RIE (SF$_6$ and O$_2$ plasma, 120 W) was done for 2 min to etch SiO$_2$ and to get hole structure of depth ~194 nm (supplementary information figure 2a) on SiO$_2$/Si.

In another case, to get the pillar structures on SiO$_2$/Si, Aluminium (Al) was sputtered (30 nm) on a circular array patterned SiO$_2$/Si after the first step of photolithography. Then deep RIE was carried out for 2 min to get Al capped SiO$_2$/Si pillar structure, where sputtered Al was used as the mask to obtain the circular pillars. Finally, the Al mask was etched using Al etchant to obtain SiO$_2$/Si pillar structures of height 186 nm (supplementary figure 2b).

### 4.3 Preparation of hBN/ WS$_2$ heterostructure

The as-grown SiO$_2$/Si substrate containing monolayer WS$_2$ was spin-coated (3500 rpm, 1 minute) with 10% polystyrene (PS) solution in chloroform [53] and was heated at 100°C for 3 minutes. The PS coated substrate was then dipped in distilled (DI) water with the four corners slightly scratched using a clean surgical blade. On dipping in DI water, the PS film comes off easily along with the monolayer WS$_2$ and the film floats in the DI water. hBN was exfoliated using scotch tape on an IPA/acetone cleaned and O$_2$ plasma-treated 300 nm SiO$_2$/ Si substrate. This hBN exfoliated substrate was then used to fish the PS film floating in water (since the WS$_2$ monolayers were on the surface of PS film which is facing the water). This substrate was then heated at 80°C for 1 hour and 150°C for 30 minutes for better adhesion of PS film with the substrate. The substrate was then dipped in toluene for 3 hours which removed the PS and was then dipped in acetone and IPA for 10 minutes in succession to remove any organic adsorbates. The substrate was then blow-dried and heated at 110°C for 10 minutes to remove water molecules. For transferring monolayer WS$_2$ on hole and pillar SiO$_2$/Si substrate, the patterns (hole and pillar) were made covering a large area such that the region of the substrate where the pillars/holes are located is easily visible to naked eyes. The floating PS film containing the monolayer WS$_2$ (on the surface facing the water) is scooped by a clean tweezer and is placed in the same orientation on the marked area of pillar/hole substrates. Removal of PS and other cleaning processes were the same as discussed earlier. For placing hBN flake on top of monolayer WS$_2$, a dry transfer technique involving polydimethylsiloxane (PDMS) stamp and a home-built transfer stage was used. PDMS stamp was made by mixing PDMS liquid and the hardener in 10:1 ratio and was subsequently cured at 90°C for 30 minutes. This PDMS stamp was then attached to a clean glass slide by plasma bonding (O$_2$ plasma, 80 W, 40 seconds). hBN was exfoliated by scotch tape on the top surface of the PDMS stamp and a suitable flake was identified by using an optical microscope for transfer. The transfer steps mentioned here [53] were followed with the stage of the transfer setup at 120°C to remove blisters at the WS$_2$/hBN interface formed during the transfer process. The sample was then dipped in acetone and IPA in succession for 10 minutes each to remove any polymer. The substrate was then blow-dried and heated at 110°C for 10 minutes to remove water molecules.

### 4.4 PL and Raman spectroscopy

The optical measurements were performed using an upright microscope (Olympus). The sample was illuminated using a 100x 0.95 NA lens at wavelength 532 nm laser excitation. The emitted light from the sample upon excitation was collected using the same lens and was projected onto the spectrometer/EMCCD for spectral analysis/imaging using relay optics. Whereas a focused excitation was used for the spectral analysis, a broad illumination configuration was employed for the PL imaging. This was done by placing a lens before the objective lens back aperture at the input path. Combination of half wave plate and polarizing beam splitter were used for varying the input excitation laser power and 532 Edge and 532 Notch filters were used to efficiently reject the Rayleigh scattered light in the output path. The Raman spectra measurements were performed with 532 nm as excitation wavelength and dispersing the signal using a 1800 lines/mm grating, keeping the centre wavelength at 540 nm. This allows us a wavelength resolution of ~0.034 nm, and consequently, ~1.15 cm$^{-1}$ resolution in wavenumber. Schematic of the optical setup (supplementary information figure 14) and the details of the excitation power measurement process is included in the supplementary information.

## Supplementary Information

Available online at free of cost.

## Author Contribution
T.C. and D.P. has contributed equally to this work.

## Conflict of interests
Authors declare no competing financial interests.

## Acknowledgements
T.C. acknowledges Prime Minister's Research Fellowship, Government of India for research fellowship. G.M.A. acknowledges UGC, Government of India for research fellowship. A.R. and D.N. acknowledge funding support from the Indo-French Centre for the Promotion of Advanced Research (CEFIPRA), project no. 6104-2. A.R. acknowledges partial funding support from DST SERB Grant no. EMR/2016/005792. T.T. and K.W. acknowledge funding for the growth of hexagonal boron nitride crystals which was supported by the Elemental Strategy Initiative conducted by the MEXT, Japan, Grant Number JPMXP0112101001 and JSPS KAKENHI Grant Number JP20H00354. D.P. thanks Sunny Tiwari for his help in alignment of the PL and Raman setup.